\documentclass[%
  preprint,
 showpacs,
 showkeys,
 preprintnumbers,
 amsmath,amssymb,
 aps,
  pra,
  longbibliography,
 ]{revtex4-1}

\usepackage{amssymb,amsthm,amsmath}

\usepackage{tikz}
\usepackage[breaklinks=true,colorlinks=true,anchorcolor=blue,citecolor=blue,filecolor=blue,menucolor=blue,pagecolor=blue,urlcolor=blue,linkcolor=blue]{hyperref}
\usepackage{graphicx}
\usepackage{url}

\usepackage{xcolor}

\begin{document}

\title{Multi-neuronal auditory coding for frequency resolution beyond the refractory threshold}

\author{Klaus Ehrenberger}
\affiliation{Emeritus, ENT Department, Medical University Vienna,
Waehringer Guertel 18-20, A-1090 Vienna, Austria}
\email{klaus.ehrenberger@meduniwien.ac.at}

\author{Karl Svozil}
\affiliation{Institute of Theoretical Physics, Vienna
    University of Technology, Wiedner Hauptstra\ss e 8-10/136, A-1040
    Vienna, Austria}
\email{svozil@tuwien.ac.at} \homepage[]{http://tph.tuwien.ac.at/~svozil}

\date{\today}

\begin{abstract}
We propose a new mechanism for high-pitch perception by a system of multiple neurons capable of resolving frequencies higher than the frequency associated with the mean refractory period up to a multiple thereof.
\end{abstract}


\maketitle

The cochlea works as a transducer of minor fluctuations in the atmospheric pressure (``sound'') into a train of action potentials along the auditory nerve. The properties of sound are represented in spatial (e.g., tonotopic) and temporal patterns of neuronal spike trains.

The spatial or place theory states that the inner ear acts as a tuned resonator. The temporal frequency theory assumes that the complete time domain representation of the incoming sound waves is directly encoded in the firing patterns of the auditory neurons.

There is still controversy about significance and interrelationship of both coding strategies~\cite{grothe-2000,oxenham-2004,Oxenham01122008,Oxenham26092012}.

In the mammalian cochlea the ribbon synapses between the inner hair cells and their afferent neurons guarantee sudden release of the neurotransmitter glutamate, triggering an irregular and bursting mode of spiking discharge. These kinetics reflect molecular instabilities of  the afferent glutamate receptors determining the mode of signal transmission in the auditory periphery~\cite{ehrenb-svozil}.

This random fractal geometry of spiking discharge patterns are processed by diverging and converging information networks of  the auditory system. In this network, the number of auditory neurons determines crucially the reliability of the auditory information flow~\cite{svoz-ehr}.

Auditory nerve fibers respond to simple acoustic stimuli with two different respond profiles~\cite{Ruggero29061992,trussel-2002,Oxenham01122008}.

For low-frequency stimuli, nerve fibers fire action potentials or spikes in a phase-locked manner, that is, with a spike occurring with a certain phase relationship to the sound stimuli.

Phase locking degrades and degenerates above circa 2 kHz. For high frequency stimuli, responses show an initial peak at the onset of the sound, and then rapid decline in firing rate down to a steady-state level of random, not phase-locked activity.

The neuronal membrane is refractory immediately after a spike, so that this firing probability, following high frequency stimulation, reflects refractoriness after preceding spike trains.

It is often argued that the limits of psychophysical performance originate in specific
stochastic neural responses, in particular, at high frequencies~\cite{heinz-2001,Oxenham26092012}.
In what follows, we present a novel mechanism utilizing stochasticity
for the transduction of sound into neural signals by considering the correlated
effect of such signals on groups of neurons, rather than considering the spike activity resulting from
a single auditory neuron.
We consider several neurons whose refractory phases are not exactly identical but vary stochastically.
Initially, the offsets of the spiking activity of these neurons also vary stochastically.
The mechanism is based on temporal single neuronal response and thereby suggest a non-tonotopic contribution to the perception of sound even beyond the refractory time.

For the sake of demonstration,
suppose these neurons are confronted with a mono-frequency signal whose pitch would require an
effective absolute refractory period of $r/n$, where $r$ is the mean refractory period
of a single neuron, and $n$ is the number of such neurons.
We will show that through the coherent stimulation of neurons,
a collective pattern of neural activity forms
which would properly contain the frequency information of the signal otherwise unattainable by single
auditory neurons.

To obtain a first feeling for this mechanism, consider a signal whose frequency is $1/r$, identical to the associated refractory period $r$ of a single neuron.
Ideally, in such a case, the temporal resolution renders the neuron to immediately fire after each refractory period.
That is,
the signal of frequency $1/r$ gets temporally resolved as $1/r$ spikes per second.

Now suppose that the frequency of the input signal is doubled, or multiplied.
In such a case, with only one neuron, this signal still is temporally resolved by merely $1/r$ spikes per second.
However, if multiple neurons are involved, multiple wave crests could activate different neurons of the group,
thereby contributing to a higher spiking activity.
For instance, if we add all spiking activity of a group of $k$ neurons, the resulting activity could result in
$k/r$ spikes per second.
In this way, the magnitude of the spiking activity is directly proportional to the frequency {\em even beyond the refractory threshold.}.

We explicitly demonstrate the aforementioned effect by a schematic, elementary model of $n=3$ neurons, all having the same
absolute refractory period $r$, which are equidistributed over $n$ periods of
length $r/n$, starting from time $t=0$.
That is, these three neurons can be successively stimulated at times
$
0,
{r\over 3},
{2r\over 3}$,
and then over again with a total offset of the
absolute refractory period $r$ of each single one of these three neurons; that is, at times
$
r,
r+{r\over 3},
r+{2r\over 3}$, and so on.

\begin{figure}
\begin{center}

\unitlength 0.200mm
\linethickness{0.4pt}
\begin{picture}(480.00,329.33)
\multiput(0.00,40.00)(0.11,0.24){12}{\line(0,1){0.24}}
\multiput(1.33,42.92)(0.11,0.22){12}{\line(0,1){0.22}}
\multiput(2.69,45.60)(0.12,0.20){12}{\line(0,1){0.20}}
\multiput(4.08,48.06)(0.12,0.19){12}{\line(0,1){0.19}}
\multiput(5.50,50.28)(0.11,0.15){13}{\line(0,1){0.15}}
\multiput(6.95,52.28)(0.11,0.14){13}{\line(0,1){0.14}}
\multiput(8.43,54.05)(0.12,0.12){13}{\line(0,1){0.12}}
\multiput(9.94,55.59)(0.14,0.12){11}{\line(1,0){0.14}}
\multiput(11.49,56.90)(0.17,0.12){9}{\line(1,0){0.17}}
\multiput(13.06,57.98)(0.20,0.11){8}{\line(1,0){0.20}}
\multiput(14.66,58.82)(0.27,0.10){6}{\line(1,0){0.27}}
\multiput(16.30,59.44)(0.42,0.10){4}{\line(1,0){0.42}}
\multiput(17.96,59.83)(1.02,0.08){2}{\line(1,0){1.02}}
\multiput(40.00,40.00)(-0.11,0.24){12}{\line(0,1){0.24}}
\multiput(38.67,42.92)(-0.11,0.22){12}{\line(0,1){0.22}}
\multiput(37.31,45.60)(-0.12,0.20){12}{\line(0,1){0.20}}
\multiput(35.92,48.06)(-0.12,0.19){12}{\line(0,1){0.19}}
\multiput(34.50,50.28)(-0.11,0.15){13}{\line(0,1){0.15}}
\multiput(33.05,52.28)(-0.11,0.14){13}{\line(0,1){0.14}}
\multiput(31.57,54.05)(-0.12,0.12){13}{\line(0,1){0.12}}
\multiput(30.06,55.59)(-0.14,0.12){11}{\line(-1,0){0.14}}
\multiput(28.51,56.90)(-0.17,0.12){9}{\line(-1,0){0.17}}
\multiput(26.94,57.98)(-0.20,0.11){8}{\line(-1,0){0.20}}
\multiput(25.34,58.82)(-0.27,0.10){6}{\line(-1,0){0.27}}
\multiput(23.70,59.44)(-0.42,0.10){4}{\line(-1,0){0.42}}
\multiput(22.04,59.83)(-1.02,0.08){2}{\line(-1,0){1.02}}
\multiput(40.00,40.00)(0.11,-0.24){12}{\line(0,-1){0.24}}
\multiput(41.33,37.08)(0.11,-0.22){12}{\line(0,-1){0.22}}
\multiput(42.69,34.40)(0.12,-0.20){12}{\line(0,-1){0.20}}
\multiput(44.08,31.94)(0.12,-0.19){12}{\line(0,-1){0.19}}
\multiput(45.50,29.72)(0.11,-0.15){13}{\line(0,-1){0.15}}
\multiput(46.95,27.72)(0.11,-0.14){13}{\line(0,-1){0.14}}
\multiput(48.43,25.95)(0.12,-0.12){13}{\line(0,-1){0.12}}
\multiput(49.94,24.41)(0.14,-0.12){11}{\line(1,0){0.14}}
\multiput(51.49,23.10)(0.17,-0.12){9}{\line(1,0){0.17}}
\multiput(53.06,22.02)(0.20,-0.11){8}{\line(1,0){0.20}}
\multiput(54.66,21.18)(0.27,-0.10){6}{\line(1,0){0.27}}
\multiput(56.30,20.56)(0.42,-0.10){4}{\line(1,0){0.42}}
\multiput(57.96,20.17)(1.02,-0.08){2}{\line(1,0){1.02}}
\multiput(80.00,40.00)(-0.11,-0.24){12}{\line(0,-1){0.24}}
\multiput(78.67,37.08)(-0.11,-0.22){12}{\line(0,-1){0.22}}
\multiput(77.31,34.40)(-0.12,-0.20){12}{\line(0,-1){0.20}}
\multiput(75.92,31.94)(-0.12,-0.19){12}{\line(0,-1){0.19}}
\multiput(74.50,29.72)(-0.11,-0.15){13}{\line(0,-1){0.15}}
\multiput(73.05,27.72)(-0.11,-0.14){13}{\line(0,-1){0.14}}
\multiput(71.57,25.95)(-0.12,-0.12){13}{\line(0,-1){0.12}}
\multiput(70.06,24.41)(-0.14,-0.12){11}{\line(-1,0){0.14}}
\multiput(68.51,23.10)(-0.17,-0.12){9}{\line(-1,0){0.17}}
\multiput(66.94,22.02)(-0.20,-0.11){8}{\line(-1,0){0.20}}
\multiput(65.34,21.18)(-0.27,-0.10){6}{\line(-1,0){0.27}}
\multiput(63.70,20.56)(-0.42,-0.10){4}{\line(-1,0){0.42}}
\multiput(62.04,20.17)(-1.02,-0.08){2}{\line(-1,0){1.02}}
\multiput(80.00,40.00)(0.11,0.24){12}{\line(0,1){0.24}}
\multiput(81.33,42.92)(0.11,0.22){12}{\line(0,1){0.22}}
\multiput(82.69,45.60)(0.12,0.20){12}{\line(0,1){0.20}}
\multiput(84.08,48.06)(0.12,0.19){12}{\line(0,1){0.19}}
\multiput(85.50,50.28)(0.11,0.15){13}{\line(0,1){0.15}}
\multiput(86.95,52.28)(0.11,0.14){13}{\line(0,1){0.14}}
\multiput(88.43,54.05)(0.12,0.12){13}{\line(0,1){0.12}}
\multiput(89.94,55.59)(0.14,0.12){11}{\line(1,0){0.14}}
\multiput(91.49,56.90)(0.17,0.12){9}{\line(1,0){0.17}}
\multiput(93.06,57.98)(0.20,0.11){8}{\line(1,0){0.20}}
\multiput(94.66,58.82)(0.27,0.10){6}{\line(1,0){0.27}}
\multiput(96.30,59.44)(0.42,0.10){4}{\line(1,0){0.42}}
\multiput(97.96,59.83)(1.02,0.08){2}{\line(1,0){1.02}}
\multiput(160.00,40.00)(0.11,0.24){12}{\line(0,1){0.24}}
\multiput(161.33,42.92)(0.11,0.22){12}{\line(0,1){0.22}}
\multiput(162.69,45.60)(0.12,0.20){12}{\line(0,1){0.20}}
\multiput(164.08,48.06)(0.12,0.19){12}{\line(0,1){0.19}}
\multiput(165.50,50.28)(0.11,0.15){13}{\line(0,1){0.15}}
\multiput(166.95,52.28)(0.11,0.14){13}{\line(0,1){0.14}}
\multiput(168.43,54.05)(0.12,0.12){13}{\line(0,1){0.12}}
\multiput(169.94,55.59)(0.14,0.12){11}{\line(1,0){0.14}}
\multiput(171.49,56.90)(0.17,0.12){9}{\line(1,0){0.17}}
\multiput(173.06,57.98)(0.20,0.11){8}{\line(1,0){0.20}}
\multiput(174.66,58.82)(0.27,0.10){6}{\line(1,0){0.27}}
\multiput(176.30,59.44)(0.42,0.10){4}{\line(1,0){0.42}}
\multiput(177.96,59.83)(1.02,0.08){2}{\line(1,0){1.02}}
\multiput(240.00,40.00)(0.11,0.24){12}{\line(0,1){0.24}}
\multiput(241.33,42.92)(0.11,0.22){12}{\line(0,1){0.22}}
\multiput(242.69,45.60)(0.12,0.20){12}{\line(0,1){0.20}}
\multiput(244.08,48.06)(0.12,0.19){12}{\line(0,1){0.19}}
\multiput(245.50,50.28)(0.11,0.15){13}{\line(0,1){0.15}}
\multiput(246.95,52.28)(0.11,0.14){13}{\line(0,1){0.14}}
\multiput(248.43,54.05)(0.12,0.12){13}{\line(0,1){0.12}}
\multiput(249.94,55.59)(0.14,0.12){11}{\line(1,0){0.14}}
\multiput(251.49,56.90)(0.17,0.12){9}{\line(1,0){0.17}}
\multiput(253.06,57.98)(0.20,0.11){8}{\line(1,0){0.20}}
\multiput(254.66,58.82)(0.27,0.10){6}{\line(1,0){0.27}}
\multiput(256.30,59.44)(0.42,0.10){4}{\line(1,0){0.42}}
\multiput(257.96,59.83)(1.02,0.08){2}{\line(1,0){1.02}}
\multiput(320.00,40.00)(0.11,0.24){12}{\line(0,1){0.24}}
\multiput(321.33,42.92)(0.11,0.22){12}{\line(0,1){0.22}}
\multiput(322.69,45.60)(0.12,0.20){12}{\line(0,1){0.20}}
\multiput(324.08,48.06)(0.12,0.19){12}{\line(0,1){0.19}}
\multiput(325.50,50.28)(0.11,0.15){13}{\line(0,1){0.15}}
\multiput(326.95,52.28)(0.11,0.14){13}{\line(0,1){0.14}}
\multiput(328.43,54.05)(0.12,0.12){13}{\line(0,1){0.12}}
\multiput(329.94,55.59)(0.14,0.12){11}{\line(1,0){0.14}}
\multiput(331.49,56.90)(0.17,0.12){9}{\line(1,0){0.17}}
\multiput(333.06,57.98)(0.20,0.11){8}{\line(1,0){0.20}}
\multiput(334.66,58.82)(0.27,0.10){6}{\line(1,0){0.27}}
\multiput(336.30,59.44)(0.42,0.10){4}{\line(1,0){0.42}}
\multiput(337.96,59.83)(1.02,0.08){2}{\line(1,0){1.02}}
\multiput(400.00,40.00)(0.11,0.24){12}{\line(0,1){0.24}}
\multiput(401.33,42.92)(0.11,0.22){12}{\line(0,1){0.22}}
\multiput(402.69,45.60)(0.12,0.20){12}{\line(0,1){0.20}}
\multiput(404.08,48.06)(0.12,0.19){12}{\line(0,1){0.19}}
\multiput(405.50,50.28)(0.11,0.15){13}{\line(0,1){0.15}}
\multiput(406.95,52.28)(0.11,0.14){13}{\line(0,1){0.14}}
\multiput(408.43,54.05)(0.12,0.12){13}{\line(0,1){0.12}}
\multiput(409.94,55.59)(0.14,0.12){11}{\line(1,0){0.14}}
\multiput(411.49,56.90)(0.17,0.12){9}{\line(1,0){0.17}}
\multiput(413.06,57.98)(0.20,0.11){8}{\line(1,0){0.20}}
\multiput(414.66,58.82)(0.27,0.10){6}{\line(1,0){0.27}}
\multiput(416.30,59.44)(0.42,0.10){4}{\line(1,0){0.42}}
\multiput(417.96,59.83)(1.02,0.08){2}{\line(1,0){1.02}}
\multiput(120.00,40.00)(-0.11,0.24){12}{\line(0,1){0.24}}
\multiput(118.67,42.92)(-0.11,0.22){12}{\line(0,1){0.22}}
\multiput(117.31,45.60)(-0.12,0.20){12}{\line(0,1){0.20}}
\multiput(115.92,48.06)(-0.12,0.19){12}{\line(0,1){0.19}}
\multiput(114.50,50.28)(-0.11,0.15){13}{\line(0,1){0.15}}
\multiput(113.05,52.28)(-0.11,0.14){13}{\line(0,1){0.14}}
\multiput(111.57,54.05)(-0.12,0.12){13}{\line(0,1){0.12}}
\multiput(110.06,55.59)(-0.14,0.12){11}{\line(-1,0){0.14}}
\multiput(108.51,56.90)(-0.17,0.12){9}{\line(-1,0){0.17}}
\multiput(106.94,57.98)(-0.20,0.11){8}{\line(-1,0){0.20}}
\multiput(105.34,58.82)(-0.27,0.10){6}{\line(-1,0){0.27}}
\multiput(103.70,59.44)(-0.42,0.10){4}{\line(-1,0){0.42}}
\multiput(102.04,59.83)(-1.02,0.08){2}{\line(-1,0){1.02}}
\multiput(200.00,40.00)(-0.11,0.24){12}{\line(0,1){0.24}}
\multiput(198.67,42.92)(-0.11,0.22){12}{\line(0,1){0.22}}
\multiput(197.31,45.60)(-0.12,0.20){12}{\line(0,1){0.20}}
\multiput(195.92,48.06)(-0.12,0.19){12}{\line(0,1){0.19}}
\multiput(194.50,50.28)(-0.11,0.15){13}{\line(0,1){0.15}}
\multiput(193.05,52.28)(-0.11,0.14){13}{\line(0,1){0.14}}
\multiput(191.57,54.05)(-0.12,0.12){13}{\line(0,1){0.12}}
\multiput(190.06,55.59)(-0.14,0.12){11}{\line(-1,0){0.14}}
\multiput(188.51,56.90)(-0.17,0.12){9}{\line(-1,0){0.17}}
\multiput(186.94,57.98)(-0.20,0.11){8}{\line(-1,0){0.20}}
\multiput(185.34,58.82)(-0.27,0.10){6}{\line(-1,0){0.27}}
\multiput(183.70,59.44)(-0.42,0.10){4}{\line(-1,0){0.42}}
\multiput(182.04,59.83)(-1.02,0.08){2}{\line(-1,0){1.02}}
\multiput(280.00,40.00)(-0.11,0.24){12}{\line(0,1){0.24}}
\multiput(278.67,42.92)(-0.11,0.22){12}{\line(0,1){0.22}}
\multiput(277.31,45.60)(-0.12,0.20){12}{\line(0,1){0.20}}
\multiput(275.92,48.06)(-0.12,0.19){12}{\line(0,1){0.19}}
\multiput(274.50,50.28)(-0.11,0.15){13}{\line(0,1){0.15}}
\multiput(273.05,52.28)(-0.11,0.14){13}{\line(0,1){0.14}}
\multiput(271.57,54.05)(-0.12,0.12){13}{\line(0,1){0.12}}
\multiput(270.06,55.59)(-0.14,0.12){11}{\line(-1,0){0.14}}
\multiput(268.51,56.90)(-0.17,0.12){9}{\line(-1,0){0.17}}
\multiput(266.94,57.98)(-0.20,0.11){8}{\line(-1,0){0.20}}
\multiput(265.34,58.82)(-0.27,0.10){6}{\line(-1,0){0.27}}
\multiput(263.70,59.44)(-0.42,0.10){4}{\line(-1,0){0.42}}
\multiput(262.04,59.83)(-1.02,0.08){2}{\line(-1,0){1.02}}
\multiput(360.00,40.00)(-0.11,0.24){12}{\line(0,1){0.24}}
\multiput(358.67,42.92)(-0.11,0.22){12}{\line(0,1){0.22}}
\multiput(357.31,45.60)(-0.12,0.20){12}{\line(0,1){0.20}}
\multiput(355.92,48.06)(-0.12,0.19){12}{\line(0,1){0.19}}
\multiput(354.50,50.28)(-0.11,0.15){13}{\line(0,1){0.15}}
\multiput(353.05,52.28)(-0.11,0.14){13}{\line(0,1){0.14}}
\multiput(351.57,54.05)(-0.12,0.12){13}{\line(0,1){0.12}}
\multiput(350.06,55.59)(-0.14,0.12){11}{\line(-1,0){0.14}}
\multiput(348.51,56.90)(-0.17,0.12){9}{\line(-1,0){0.17}}
\multiput(346.94,57.98)(-0.20,0.11){8}{\line(-1,0){0.20}}
\multiput(345.34,58.82)(-0.27,0.10){6}{\line(-1,0){0.27}}
\multiput(343.70,59.44)(-0.42,0.10){4}{\line(-1,0){0.42}}
\multiput(342.04,59.83)(-1.02,0.08){2}{\line(-1,0){1.02}}
\multiput(440.00,40.00)(-0.11,0.24){12}{\line(0,1){0.24}}
\multiput(438.67,42.92)(-0.11,0.22){12}{\line(0,1){0.22}}
\multiput(437.31,45.60)(-0.12,0.20){12}{\line(0,1){0.20}}
\multiput(435.92,48.06)(-0.12,0.19){12}{\line(0,1){0.19}}
\multiput(434.50,50.28)(-0.11,0.15){13}{\line(0,1){0.15}}
\multiput(433.05,52.28)(-0.11,0.14){13}{\line(0,1){0.14}}
\multiput(431.57,54.05)(-0.12,0.12){13}{\line(0,1){0.12}}
\multiput(430.06,55.59)(-0.14,0.12){11}{\line(-1,0){0.14}}
\multiput(428.51,56.90)(-0.17,0.12){9}{\line(-1,0){0.17}}
\multiput(426.94,57.98)(-0.20,0.11){8}{\line(-1,0){0.20}}
\multiput(425.34,58.82)(-0.27,0.10){6}{\line(-1,0){0.27}}
\multiput(423.70,59.44)(-0.42,0.10){4}{\line(-1,0){0.42}}
\multiput(422.04,59.83)(-1.02,0.08){2}{\line(-1,0){1.02}}
\multiput(120.00,40.00)(0.11,-0.24){12}{\line(0,-1){0.24}}
\multiput(121.33,37.08)(0.11,-0.22){12}{\line(0,-1){0.22}}
\multiput(122.69,34.40)(0.12,-0.20){12}{\line(0,-1){0.20}}
\multiput(124.08,31.94)(0.12,-0.19){12}{\line(0,-1){0.19}}
\multiput(125.50,29.72)(0.11,-0.15){13}{\line(0,-1){0.15}}
\multiput(126.95,27.72)(0.11,-0.14){13}{\line(0,-1){0.14}}
\multiput(128.43,25.95)(0.12,-0.12){13}{\line(0,-1){0.12}}
\multiput(129.94,24.41)(0.14,-0.12){11}{\line(1,0){0.14}}
\multiput(131.49,23.10)(0.17,-0.12){9}{\line(1,0){0.17}}
\multiput(133.06,22.02)(0.20,-0.11){8}{\line(1,0){0.20}}
\multiput(134.66,21.18)(0.27,-0.10){6}{\line(1,0){0.27}}
\multiput(136.30,20.56)(0.42,-0.10){4}{\line(1,0){0.42}}
\multiput(137.96,20.17)(1.02,-0.08){2}{\line(1,0){1.02}}
\multiput(200.00,40.00)(0.11,-0.24){12}{\line(0,-1){0.24}}
\multiput(201.33,37.08)(0.11,-0.22){12}{\line(0,-1){0.22}}
\multiput(202.69,34.40)(0.12,-0.20){12}{\line(0,-1){0.20}}
\multiput(204.08,31.94)(0.12,-0.19){12}{\line(0,-1){0.19}}
\multiput(205.50,29.72)(0.11,-0.15){13}{\line(0,-1){0.15}}
\multiput(206.95,27.72)(0.11,-0.14){13}{\line(0,-1){0.14}}
\multiput(208.43,25.95)(0.12,-0.12){13}{\line(0,-1){0.12}}
\multiput(209.94,24.41)(0.14,-0.12){11}{\line(1,0){0.14}}
\multiput(211.49,23.10)(0.17,-0.12){9}{\line(1,0){0.17}}
\multiput(213.06,22.02)(0.20,-0.11){8}{\line(1,0){0.20}}
\multiput(214.66,21.18)(0.27,-0.10){6}{\line(1,0){0.27}}
\multiput(216.30,20.56)(0.42,-0.10){4}{\line(1,0){0.42}}
\multiput(217.96,20.17)(1.02,-0.08){2}{\line(1,0){1.02}}
\multiput(280.00,40.00)(0.11,-0.24){12}{\line(0,-1){0.24}}
\multiput(281.33,37.08)(0.11,-0.22){12}{\line(0,-1){0.22}}
\multiput(282.69,34.40)(0.12,-0.20){12}{\line(0,-1){0.20}}
\multiput(284.08,31.94)(0.12,-0.19){12}{\line(0,-1){0.19}}
\multiput(285.50,29.72)(0.11,-0.15){13}{\line(0,-1){0.15}}
\multiput(286.95,27.72)(0.11,-0.14){13}{\line(0,-1){0.14}}
\multiput(288.43,25.95)(0.12,-0.12){13}{\line(0,-1){0.12}}
\multiput(289.94,24.41)(0.14,-0.12){11}{\line(1,0){0.14}}
\multiput(291.49,23.10)(0.17,-0.12){9}{\line(1,0){0.17}}
\multiput(293.06,22.02)(0.20,-0.11){8}{\line(1,0){0.20}}
\multiput(294.66,21.18)(0.27,-0.10){6}{\line(1,0){0.27}}
\multiput(296.30,20.56)(0.42,-0.10){4}{\line(1,0){0.42}}
\multiput(297.96,20.17)(1.02,-0.08){2}{\line(1,0){1.02}}
\multiput(360.00,40.00)(0.11,-0.24){12}{\line(0,-1){0.24}}
\multiput(361.33,37.08)(0.11,-0.22){12}{\line(0,-1){0.22}}
\multiput(362.69,34.40)(0.12,-0.20){12}{\line(0,-1){0.20}}
\multiput(364.08,31.94)(0.12,-0.19){12}{\line(0,-1){0.19}}
\multiput(365.50,29.72)(0.11,-0.15){13}{\line(0,-1){0.15}}
\multiput(366.95,27.72)(0.11,-0.14){13}{\line(0,-1){0.14}}
\multiput(368.43,25.95)(0.12,-0.12){13}{\line(0,-1){0.12}}
\multiput(369.94,24.41)(0.14,-0.12){11}{\line(1,0){0.14}}
\multiput(371.49,23.10)(0.17,-0.12){9}{\line(1,0){0.17}}
\multiput(373.06,22.02)(0.20,-0.11){8}{\line(1,0){0.20}}
\multiput(374.66,21.18)(0.27,-0.10){6}{\line(1,0){0.27}}
\multiput(376.30,20.56)(0.42,-0.10){4}{\line(1,0){0.42}}
\multiput(377.96,20.17)(1.02,-0.08){2}{\line(1,0){1.02}}
\multiput(440.00,40.00)(0.11,-0.24){12}{\line(0,-1){0.24}}
\multiput(441.33,37.08)(0.11,-0.22){12}{\line(0,-1){0.22}}
\multiput(442.69,34.40)(0.12,-0.20){12}{\line(0,-1){0.20}}
\multiput(444.08,31.94)(0.12,-0.19){12}{\line(0,-1){0.19}}
\multiput(445.50,29.72)(0.11,-0.15){13}{\line(0,-1){0.15}}
\multiput(446.95,27.72)(0.11,-0.14){13}{\line(0,-1){0.14}}
\multiput(448.43,25.95)(0.12,-0.12){13}{\line(0,-1){0.12}}
\multiput(449.94,24.41)(0.14,-0.12){11}{\line(1,0){0.14}}
\multiput(451.49,23.10)(0.17,-0.12){9}{\line(1,0){0.17}}
\multiput(453.06,22.02)(0.20,-0.11){8}{\line(1,0){0.20}}
\multiput(454.66,21.18)(0.27,-0.10){6}{\line(1,0){0.27}}
\multiput(456.30,20.56)(0.42,-0.10){4}{\line(1,0){0.42}}
\multiput(457.96,20.17)(1.02,-0.08){2}{\line(1,0){1.02}}
\multiput(160.00,40.00)(-0.11,-0.24){12}{\line(0,-1){0.24}}
\multiput(158.67,37.08)(-0.11,-0.22){12}{\line(0,-1){0.22}}
\multiput(157.31,34.40)(-0.12,-0.20){12}{\line(0,-1){0.20}}
\multiput(155.92,31.94)(-0.12,-0.19){12}{\line(0,-1){0.19}}
\multiput(154.50,29.72)(-0.11,-0.15){13}{\line(0,-1){0.15}}
\multiput(153.05,27.72)(-0.11,-0.14){13}{\line(0,-1){0.14}}
\multiput(151.57,25.95)(-0.12,-0.12){13}{\line(0,-1){0.12}}
\multiput(150.06,24.41)(-0.14,-0.12){11}{\line(-1,0){0.14}}
\multiput(148.51,23.10)(-0.17,-0.12){9}{\line(-1,0){0.17}}
\multiput(146.94,22.02)(-0.20,-0.11){8}{\line(-1,0){0.20}}
\multiput(145.34,21.18)(-0.27,-0.10){6}{\line(-1,0){0.27}}
\multiput(143.70,20.56)(-0.42,-0.10){4}{\line(-1,0){0.42}}
\multiput(142.04,20.17)(-1.02,-0.08){2}{\line(-1,0){1.02}}
\multiput(240.00,40.00)(-0.11,-0.24){12}{\line(0,-1){0.24}}
\multiput(238.67,37.08)(-0.11,-0.22){12}{\line(0,-1){0.22}}
\multiput(237.31,34.40)(-0.12,-0.20){12}{\line(0,-1){0.20}}
\multiput(235.92,31.94)(-0.12,-0.19){12}{\line(0,-1){0.19}}
\multiput(234.50,29.72)(-0.11,-0.15){13}{\line(0,-1){0.15}}
\multiput(233.05,27.72)(-0.11,-0.14){13}{\line(0,-1){0.14}}
\multiput(231.57,25.95)(-0.12,-0.12){13}{\line(0,-1){0.12}}
\multiput(230.06,24.41)(-0.14,-0.12){11}{\line(-1,0){0.14}}
\multiput(228.51,23.10)(-0.17,-0.12){9}{\line(-1,0){0.17}}
\multiput(226.94,22.02)(-0.20,-0.11){8}{\line(-1,0){0.20}}
\multiput(225.34,21.18)(-0.27,-0.10){6}{\line(-1,0){0.27}}
\multiput(223.70,20.56)(-0.42,-0.10){4}{\line(-1,0){0.42}}
\multiput(222.04,20.17)(-1.02,-0.08){2}{\line(-1,0){1.02}}
\multiput(320.00,40.00)(-0.11,-0.24){12}{\line(0,-1){0.24}}
\multiput(318.67,37.08)(-0.11,-0.22){12}{\line(0,-1){0.22}}
\multiput(317.31,34.40)(-0.12,-0.20){12}{\line(0,-1){0.20}}
\multiput(315.92,31.94)(-0.12,-0.19){12}{\line(0,-1){0.19}}
\multiput(314.50,29.72)(-0.11,-0.15){13}{\line(0,-1){0.15}}
\multiput(313.05,27.72)(-0.11,-0.14){13}{\line(0,-1){0.14}}
\multiput(311.57,25.95)(-0.12,-0.12){13}{\line(0,-1){0.12}}
\multiput(310.06,24.41)(-0.14,-0.12){11}{\line(-1,0){0.14}}
\multiput(308.51,23.10)(-0.17,-0.12){9}{\line(-1,0){0.17}}
\multiput(306.94,22.02)(-0.20,-0.11){8}{\line(-1,0){0.20}}
\multiput(305.34,21.18)(-0.27,-0.10){6}{\line(-1,0){0.27}}
\multiput(303.70,20.56)(-0.42,-0.10){4}{\line(-1,0){0.42}}
\multiput(302.04,20.17)(-1.02,-0.08){2}{\line(-1,0){1.02}}
\multiput(400.00,40.00)(-0.11,-0.24){12}{\line(0,-1){0.24}}
\multiput(398.67,37.08)(-0.11,-0.22){12}{\line(0,-1){0.22}}
\multiput(397.31,34.40)(-0.12,-0.20){12}{\line(0,-1){0.20}}
\multiput(395.92,31.94)(-0.12,-0.19){12}{\line(0,-1){0.19}}
\multiput(394.50,29.72)(-0.11,-0.15){13}{\line(0,-1){0.15}}
\multiput(393.05,27.72)(-0.11,-0.14){13}{\line(0,-1){0.14}}
\multiput(391.57,25.95)(-0.12,-0.12){13}{\line(0,-1){0.12}}
\multiput(390.06,24.41)(-0.14,-0.12){11}{\line(-1,0){0.14}}
\multiput(388.51,23.10)(-0.17,-0.12){9}{\line(-1,0){0.17}}
\multiput(386.94,22.02)(-0.20,-0.11){8}{\line(-1,0){0.20}}
\multiput(385.34,21.18)(-0.27,-0.10){6}{\line(-1,0){0.27}}
\multiput(383.70,20.56)(-0.42,-0.10){4}{\line(-1,0){0.42}}
\multiput(382.04,20.17)(-1.02,-0.08){2}{\line(-1,0){1.02}}
\multiput(480.00,40.00)(-0.11,-0.24){12}{\line(0,-1){0.24}}
\multiput(478.67,37.08)(-0.11,-0.22){12}{\line(0,-1){0.22}}
\multiput(477.31,34.40)(-0.12,-0.20){12}{\line(0,-1){0.20}}
\multiput(475.92,31.94)(-0.12,-0.19){12}{\line(0,-1){0.19}}
\multiput(474.50,29.72)(-0.11,-0.15){13}{\line(0,-1){0.15}}
\multiput(473.05,27.72)(-0.11,-0.14){13}{\line(0,-1){0.14}}
\multiput(471.57,25.95)(-0.12,-0.12){13}{\line(0,-1){0.12}}
\multiput(470.06,24.41)(-0.14,-0.12){11}{\line(-1,0){0.14}}
\multiput(468.51,23.10)(-0.17,-0.12){9}{\line(-1,0){0.17}}
\multiput(466.94,22.02)(-0.20,-0.11){8}{\line(-1,0){0.20}}
\multiput(465.34,21.18)(-0.27,-0.10){6}{\line(-1,0){0.27}}
\multiput(463.70,20.56)(-0.42,-0.10){4}{\line(-1,0){0.42}}
\multiput(462.04,20.17)(-1.02,-0.08){2}{\line(-1,0){1.02}}
\put(20.00,200.00){\line(1,0){240.00}}
\put(20.00,200.00){\circle*{5.20}}
\put(260.00,200.00){\circle*{5.20}}
\put(100.00,150.00){\line(1,0){240.00}}
\put(100.00,150.00){\circle*{5.20}}
\put(340.00,150.00){\circle*{5.20}}
\put(180.00,100.00){\line(1,0){240.00}}
\put(180.00,100.00){\circle*{5.20}}
\put(420.00,100.00){\circle*{5.20}}
\put(20.00,220.00){\line(0,-1){205.33}}
\put(260.00,220.00){\line(0,-1){205.33}}
\put(100.00,169.33){\line(0,-1){156.00}}
\put(340.00,169.33){\line(0,-1){156.00}}
\put(180.00,118.67){\line(0,-1){105.33}}
\put(420.00,118.67){\line(0,-1){105.33}}
\put(20.00,0.00){\makebox(0,0)[cc]{$0$}}
\put(100.00,0.00){\makebox(0,0)[cc]{$r\over 3$}}
\put(180.00,0.00){\makebox(0,0)[cc]{$2r\over 3$}}
\put(260.00,0.00){\makebox(0,0)[cc]{$r$}}
\put(340.00,0.00){\makebox(0,0)[cc]{$4r\over 3$}}
\put(420.00,0.00){\makebox(0,0)[cc]{$5r\over3$}}
\put(140.00,220.00){\makebox(0,0)[cc]{$r$}}
\put(220.00,169.33){\makebox(0,0)[cc]{$r$}}
\put(300.00,118.67){\makebox(0,0)[cc]{$r$}}
\put(282.67,200.00){\makebox(0,0)[lc]{neuron \# 1}}
\put(362.67,149.33){\makebox(0,0)[lc]{neuron \# 2}}
\put(442.67,98.67){\makebox(0,0)[lc]{neuron \# 3}}
\put(0.00,249.33){\line(1,0){480.00}}
\put(20.00,249.33){\vector(0,1){50.67}}
\put(100.00,249.33){\vector(0,1){50.67}}
\put(180.00,249.33){\vector(0,1){50.67}}
\put(260.00,249.33){\vector(0,1){50.67}}
\put(340.00,249.33){\vector(0,1){50.67}}
\put(420.00,249.33){\vector(0,1){50.67}}
\put(20.00,329.33){\makebox(0,0)[cc]{\# 1}}
\put(100.00,329.33){\makebox(0,0)[cc]{\# 2}}
\put(180.00,329.33){\makebox(0,0)[cc]{\# 3}}
\put(260.00,329.33){\makebox(0,0)[cc]{\# 1}}
\put(340.00,329.33){\makebox(0,0)[cc]{\# 2}}
\put(420.00,329.33){\makebox(0,0)[cc]{\# 3}}
\put(0.00,20.00){\makebox(0,0)[cc]{a)}}
\put(0.00,150.00){\makebox(0,0)[cc]{b)}}
\put(0.00,280.00){\makebox(0,0)[cc]{c)}}
\end{picture}
\end{center}
\caption{Temporal evolution of a system of three neurons with equidistributed
onsets of identical absolute refractory periods $r$, stimulated by
successive wave peaks and thus being capable of resolving
signals of frequency $3/r$.
a) Original signal; b) neuron activation cycle; c) sum of spike trains from neuron activity.
\label{2006-highpitch-f1}}
\end{figure}
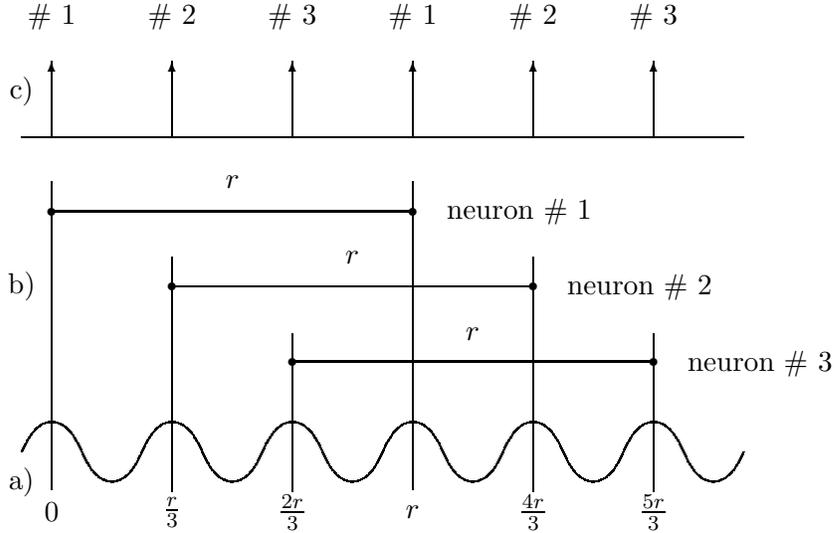

For such a configuration, each one of the neurons can take up a signal for the successive wave
crests at a frequency $3\over r$.
Fig. \ref{2006-highpitch-f1} depicts the temporal evolution of this system of neurons, stimulated by
successive wave peaks.

This effect is based on the assumption that the neurons either have a different offset of the refractory period,
or have different absolute refractory periods (within a certain frequency range).
In such cases, different neurons are stimulated by successive signal peaks.
The sum over the neural activity of this group of neurons then properly represents
information about the high-pitch signal, even if its frequency is too
high to be resolved by a single neuron alone.

The price to be paid for this ``optimal'' resolution of a mono-frequency signal is
the narrow (indeed, of width zero) bandwidth which is resolved by the three neurons.
This can be circumvented by considering a {\em stochastic} distribution of the offset phases.
Stochastic offsets will be discussed below in greater detail.

Another issue is the attenuation of the signal by an effective factor of $n$
with respect to the single, stand-alone neuron activation in the case of signals with frequencies
so low that they can be temporally resolved within the absolute refractory period.
This attenuation should be compensated by either the plasticity of the auditory perception system,
or by the integration of more neurons which effectively contribute to the overall signal.

In what follows we present detailed numerical studies of multi-neuronal systems
with a stochastic distribution of
absolute refractory periods within an interval $\left[r-{\Delta \over 2},r+{\Delta \over 2}\right]$ and
initial offsets of the order of $r$.
The driving signal is modelled by a regular spiking activity of Frequency $\omega = k / r$.
In the  $k >1$ regime, coherent stimulation can be expected to contribute to
high pitch perception.
As for the regular case described above, the mechanism can be expected to work for $k \le n$.

\begin{figure}
  \centering
\begin{tabular}{ c c c }
  \includegraphics[width=70mm]{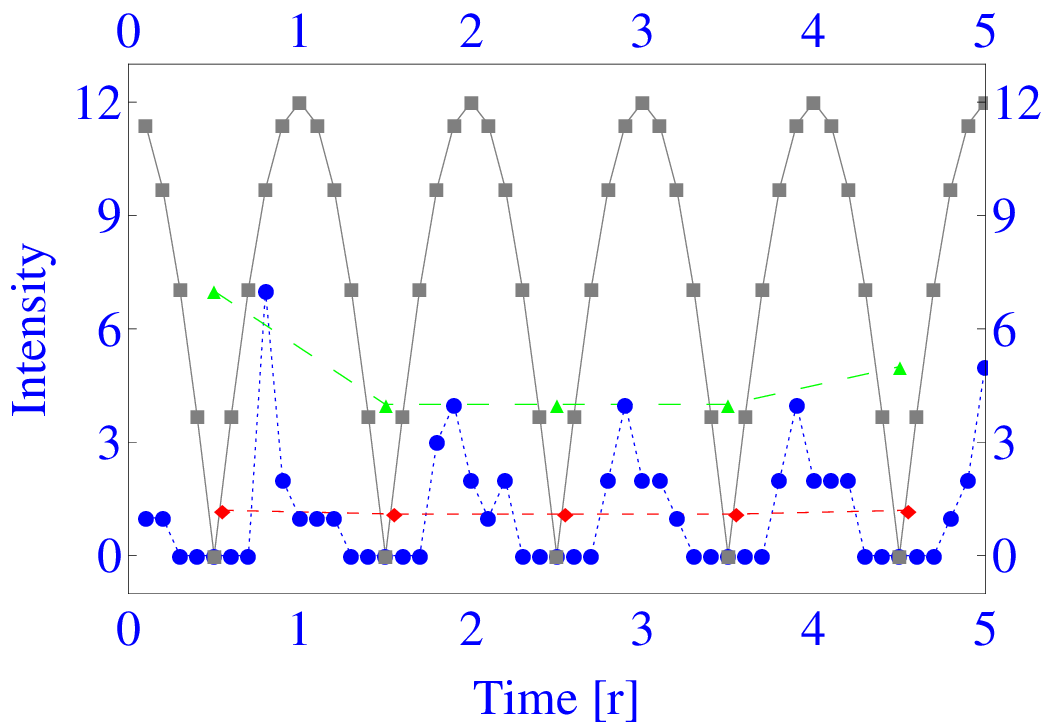} & \qquad & \includegraphics[width=70mm]{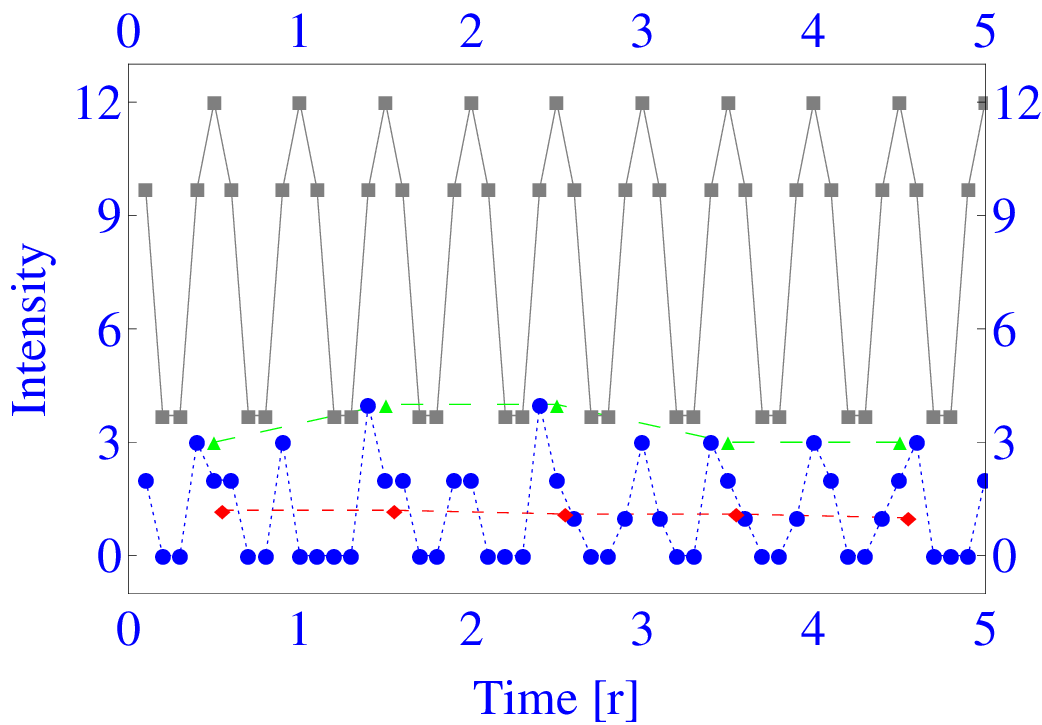}\\
(a)&&(b)\\
\end{tabular}
  \caption{\label{2014-highpitch-f2} (Color online)
Numerical simulation of the intensity of the spiking activity as a result of $12$ neurons driven by a signal
corresponding to (a) one and (b) half the inverse mean absolute refractory period.
The grey solid line with quadratic markers indicates the (discretized) signal.
The blue dotted line indicates the neuronal response.
The red small-dashed and green large-dashed lines indicate mean and maximum of the neuronal response, respectively.
}
\end{figure}

Fig.~\ref{2014-highpitch-f2} depicts a numerical simulation
of the intensity of the spiking activity as a result of $12$ neurons driven by a signal of twice frequency corresponding to twice
the inverse mean absolute refractory period.
The numerical studies indicate a reliable performance of coherent stimulation for
frequencies corresponding to lower than or equal to the number of participating neurons.
The relative spread of the refractory periods of the single neurons has been chosen to be 20\% of the refractory period.

Numerical simulations also show that, whereas the signal is rather well represented by the responses of the single neurons,
the mean and, to a lesser degree, the maximum of the neural response does not change significantly if the frequency is varied.
This is due to the statistical averaging of the spiking activity of the group of neurons interacting with the signal.

In summary, we have presented one theoretical mechanism of high-pitch sound perception
and one practical application thereof.
Theoretically,  we have demonstrated a novel mechanism of high-pitch perception for the auditory transduction of sound into neural signals.
This mechanism utilizes stochasticity in a system of multiple neurons,
whose collective excitations resolve frequencies higher than the frequency
associated with the mean refractory period~\cite{Berry15031998} up to a multiple thereof.
It may contribute to a non-tonotopic mode of high frequency perception.
As a practical application we suggest an economic solution for a single electrode cochlear implant
which may yield speech discrimination through this mechanism.

\section*{Acknowledgement}
This work was supported in part by Marie Curie FP7-PEOPLE-2010-IRSES Grant RANPHYS.


%

\end{document}